# Measuring Undergraduate Students' Reliance on Generative AI During Problem-Solving: Scale Development and Validation

This is the author's accepted manuscript version of the following article:

Hou, C., Zhu, G., Vidya Sudarshan, Lim, F. S., & Ong, Y. S. (2025). Measuring undergraduate students' reliance on generative AI during problem-solving: Scale development and validation. *Computers & Education, 234*, Article 105329. https://doi.org/10.1016/j.compedu.2025.105329

**For people who are interested in using or adapting the scale, please fill in this form:**

**https://forms.gle/TiAMgYC81ZzbWQKB9**

## Abstract

Reliance on AI describes the behavioral patterns of when and how individuals depend on AI suggestions, and appropriate reliance patterns are necessary to achieve effective human-AI collaboration. Traditional measures often link reliance to decision-making outcomes, which may not be suitable for complex problem-solving tasks where outcomes are not binary (i.e., correct or incorrect) or immediately clear. Therefore, this study aims to develop a scale to measure undergraduate students' behaviors of using Generative AI during problem-solving tasks without directly linking them to specific outcomes. We conducted an exploratory factor analysis on 800 responses collected after students finished one problem-solving activity, which revealed four distinct factors: reflective use, cautious use, thoughtless use, and collaborative use. The overall scale has reached sufficient internal reliability (Cronbach's alpha = .84). Two confirmatory factor analyses (CFAs) were conducted to validate the factors using the remaining 730 responses from this activity and 1,173 responses from another problem-solving activity. CFA indices showed adequate model fit for data from both problem-solving tasks, suggesting that the scale can be applied to various human-AI problem-solving tasks. This study offers a validated scale to measure students' reliance behaviors in different human-AI problem-solving activities and provides implications for educators to responsively integrate Generative AI in higher education.

## Introduction

Artificial intelligence (AI), defined as digital machines capable of performing cognitive tasks similar to human minds, such as learning and problem-solving, has revolutionized the ways people teach and learn (Chiu et al., 2023; Tahiru, 2021). AI, particularly Generative AI, has emerged as a tool that can support learning, offering students assistance in exploring solutions and generating ideas (Grassini, 2023; Sleator, 2023). In higher education, problem-based learning (PBL) is increasingly recognized for its emphasis on real-world and interdisciplinary challenges (Strobel & Barneveld, 2009), which can actively engage students in complex problem-solving processes, encourage them to explore diverse perspectives and rich information, and develop practical solutions through iterative inquiries (Wijnia et al., 2019). However, introducing AI in PBL poses significant challenges, particularly concerning potential students' overreliance on AI tools which could undermine their problem-solving skills. Studies have shown that students might accept automated feedback provided by AI tools with minimal cognitive engagement (Kasneci et al., 2023; Koltovskaia, 2020). Inappropriate reliance on AI may erode students' ability to critically evaluate information and engage in metacognitive practices in the PBL context (Hamid et al., 2023). Therefore, understanding how college students interact with and depend on Generative AI during PBL is crucial for ensuring that learning goals are not being compromised because of students' inappropriate reliance behaviors.

Reliance on AI systems, the behavioral patterns of when and how individuals depend on AI suggestions (Wang et al., 2008), is a pivotal aspect of human-AI collaboration that can influence whether AI systems enhance or diminish human performance. Introducing AI to assist complex tasks can enhance overall performance, but individuals need to develop the ability to identify and reject erroneous AI advice (Bansal et al., 2019). Several measures have been developed to quantify reliance behaviors, typically emphasizing collective performance between AI and humans. For instance, some instruments measure reliance based on the decision-making outcomes (e.g., Schemmer et al., 2022), while others take a prospective method to measure reliance as the probability that humans would adopt AI suggestions to maximize potential benefits (e.g., Fok & Weld, 2024). These measures are limited in their ability to capture students' inobservable reliance behaviors on Generative AI during problem-solving tasks that involve implicit shifts in cognitive strategies based on interactions with the environment or other participants (Chiou & Lee, 2023). Indeed, problem-solving is a dynamic and multifaceted activity that requires ongoing collaboration and adaptation among participants. They need to develop shared mental models by communicating and integrating diverse knowledge to achieve a common goal (Larson Jr & Christensen, 1993). This process encompasses multiple interrelated stages, such as problem identification, information exploration, and solution generation and presentation (Liu & Pásztor, 2022), each requiring cognitive skills to carry out (Behfar et al., 2008). Effective problem-solving is nonlinear and a more intricate process than decision-making, which typically involves explicit choices and outcomes (Hesse et al., 2015).

Recognizing the limitations of existing measures, the current study aims to develop a scale that measures reliance behaviors during human-AI problem-solving. As Generative AI tools are incorporated into collaborative problem-solving tasks in classrooms, there is a pressing need to assess how students make use of Generative AI's suggestions, as inappropriate reliance on AI could exacerbate inequalities in learning and compromise the rigorousness of academic works (Al-kfairy et al., 2024). The self-report questionnaire aims to address this need by providing a nuanced, user-centered measure of AI reliance in the human-AI problem-solving context. By enumerating use cases of Generative AI across the three stages (i.e., problem identification, information exploration, and solution generation and presentation; Liu & Pásztor, 2022) of problem-solving, the current study seeks to uncover patterns of reliance that might not be easily observed through objective measures. Furthermore, leveraging a large-scale survey in a course taken by undergraduates across a university, this research will provide an overview of how undergraduate students interact with Generative AI in problem-solving activities. Specifically, the research questions we aim to address are:

1) What dimensions of undergraduate students' reliance behaviors emerge from question items?
2) To what extent do the developed items demonstrate reliability across different types of problem-solving tasks involving Generative AI?
3) How do undergraduate students' reliance behaviors on Generative AI differ across problem-solving tasks?

## Literature Review

### Problem-Based Learning and Problem-Solving

PBL is an instructional approach widely adopted in higher education, focusing on student-driven collaborative exploration of real-world, ill-structured problems. These problems do not have clearly defined solutions or paths to resolution but provide real and complex scenarios for inquiry, collaboration between students, sharing of diverse perspectives, integration of ideas, and deep learning (Barrows, 1996; Wijnia et al., 2019). PBL fosters the development of procedural knowledge, such as reasoning, and declarative knowledge, which involves understanding the principles underlying a problem (Schmidt et al., 2009).

The central focus of PBL is problem-solving. While there are various models of PBL, they all tend to follow a structured problem-solving process involving three key stages: problem identification, information exploration, and solution generation and presentation (Liu & Pásztor, 2022). During problem identification, students need to define the problems within the given problem scope and brainstorm initial ideas or hypotheses to identify knowledge gaps (Schmidt, 1993, 2012). In the information exploration stage, students engage in iterative research to address these gaps, sharing and analyzing information while connecting it to broader concepts (Wijnia et al., 2019). Finally, in the solution generation and presentation stage, students consolidate their findings, develop solutions, and share their solutions with the larger community, reflecting on both the solutions and their learning process (Koschmann, 1994).

### ***Human-AI Problem-Solving***

Ideally, human-AI collaboration, where humans and AI systems work together by leveraging their unique strengths, can enhance human cognition (Dellermann et al., 2019). Specifically, AI can quickly process vast amounts of data and identify patterns, while humans can contribute nuanced reasoning, creativity, and contextual understanding (Braga & Logan, 2017). In educational settings, AI can enhance learning by offering personalized instruction, modeling learning behaviors, supporting formative assessments, and providing timely feedback (Tan et al., 2022). In problem-solving, AI technologies have evolved from simply reducing labor costs to providing cognitive support in tasks such as data analysis, pattern recognition, and even generating creative solutions (Joksimovic et al., 2023; Seeber et al., 2020). In addition, AI tools can provide feedback on the problem-solving process to help learners monitor and reflect on their problem-solving strategies (Joksimovic et al., 2023).

Furthermore, Generative AI, a category of AI that can produce diverse outputs such as text, images, music, and code based on users' natural language input (i.e., prompt engineering), is reshaping educational practices (Zhao et al., 2023). Generative AI has demonstrated impressive capabilities across complex tasks like arithmetic problems, conceptual mappings, and multidisciplinary understanding (Wei et al., 2022). Hence, Generative AI has the potential to facilitate students' problem-solving during PBL. Generative AI can aid the problem-solving process by acting as a tool for knowledge generation, idea exploration, and feedback (Joksimovic et al., 2023). In the problem identification stage, Generative AI may help brainstorm potential hypotheses and simulate different problem scenarios. For example, Valdivia and colleagues (2024) used Generative AI to generate different problem scenarios concerning unit conversion and thermodynamic processes for students to solve. Their results show that students demonstrate higher-order cognitive levels, including analyzing, synthesizing, and evaluating in these problem-solving tasks. During information exploration, Generative AI can assist in gathering relevant information, summarizing complex and rich information, and even suggesting new directions for inquiry. For example, the study on the GPT-Assisted Summarization Aid (GASA, Lin et al., 2024) demonstrated how Generative AI helped students in STEM education by summarizing group discussions, providing real-time feedback on technical questions, and guiding reflective thinking. GASA can support students in identifying and organizing key concepts during collaborative tasks and suggest advanced approaches for integrating knowledge. Lastly, in the solution generation and presentation stage, Generative AI can offer alternative perspectives, help students refine their solutions, and help them prepare coherent solutions. For example, in Naik's et al. (2024) study on collaborative programming learning, a GPT-based tool was integrated into a cloud-based platform to provide dynamic and personalized reflection triggers. These triggers, tailored to the Structured Query Language (SQL) commands entered by student groups, prompted alternative solutions and facilitated discussions on key programming concepts, such as datatype selection, indexing strategies, and denormalization trade-offs. While the integration of these tailored reflection triggers influenced how students allocated their time, changing completion rates per problem, it did not diminish the overall amount learned.

Integrating Generative AI into PBL has the potential to provide students with tailored feedback and engage them in deeper problem exploration, making the learning process more dynamic and efficient. For example, Hamid and colleagues (2023) investigated the use of ChatGPT in pharmaceutical chemistry PBL sessions with 18 participants. Qualitative analysis of the collaborative process showed that integrating GPT in the process enhanced collaboration, engagement, and motivation, but the accuracy and depth of AI-generated information required additional human verification. These findings underscore the dual-edged nature of integrating Generative AI in education. Issues such as misinformation, biases embedded in AI algorithms, and academic dishonesty compound the risks of AI integration in education (Al-kfairy et al., 2024). Generative AI tools may inadvertently reinforce stereotypes or factual inaccuracies due to biases in training data, creating inequities in learning outcomes (Hacker et al., 2023). In addition, AI generated content raises questions about authorship and intellectual integrity, necessitating transparency in AI usage and the development of mechanisms for attribution and verification (Malik et al., 2023).

Indeed, inappropriate reliance on Generative AI could undermine students' high-order competencies such as critical thinking and problem-solving skills (Kasneci et al., 2023). Research has shown that students often accept automated feedback with little cognitive engagement, highlighting the risk of thoughtless reliance on AI tools (Koltovskaia, 2020). Such passive acceptance of AI-generated responses can erode critical thinking and analytical processes central to PBL (Kadaruddin, 2023; Sun et al., 2025). Thus, it is vital for educators to actively monitor students’ interactions with Generative AI to ensure it supplements rather than supplants their cognitive effort.

Thus, more research is needed to systematically examine the use of Generative AI in PBL (Hamid et al., 2023). The first step is to develop proper measures to uncover how students use Generative AI during problem-solving tasks. The measure should evaluate not only the frequency of AI use but also the depth of students’ cognitive engagement with AI-generated content. Emphasis must be placed on distinguishing between user behaviors that support learning and detrimental overreliance that stifles independent reasoning.

**Measures for Reliance on AI Systems**

Based on the definition of reliance behaviors on AI systems, previous studies have sought to quantify these behaviors through various empirical measures (e.g., Buçinca et al., 2021; Vasconcelos et al., 2023). These measures typically focus on observable traces of reliance on AI suggestions. For instance, the switch fraction measures the ratio of users changing their answers to match AI suggestions compared to all decision-making trials (A. Kim et al., 2023). Another common measure, the Weight of Advice, evaluates how much influence AI's recommendations have on users' final decisions relative to their initial independent choices. If the final decision aligns more closely with the AI's recommendation than the user's initial judgment, it indicates a higher reliance on AI (Passi & Vorvoreanu, 2022). However, these measures fail to differentiate between correct and incorrect AI suggestions, neglecting the fact that AI can also make errors. Users need to discern the quality of AI advice and adjust their reliance accordingly.

To address the limitations of the measures, Schemmer et al. (2022) proposed that measures for reliance on AI should also emphasize users' ability to discriminate between correct and incorrect AI advice and act accordingly. Their measure involves a sequential decisionmaking process where users make an initial decision, receive AI advice, and then decide whether to follow or reject the advice. This model introduces two dimensions: relative AI reliance, which tracks how often users rightfully follow AI recommendations, and relative self-reliance, which measures how often users correctly reject incorrect AI advice. Although this approach provides a more nuanced understanding of reliance, it remains limited to binary decision-making contexts. This measure relies heavily on immediate feedback about the correctness of decisions, which is not always available in real-world settings. For example, in financial decisions like investments, the appropriateness of relying on AI predictions may fluctuate over time, depending on market conditions. Thus, the two-dimensional reliance construct is restricted to structured, outcomedependent decision-making tasks and does not account for more complex, non-binary human-AI collaboration scenarios (Fok & Weld, 2024).

To overcome the above-mentioned limitations, Fok and Weld (2024) introduced the concept of strategy-graded reliance, which focuses on reliance behaviors based on expected AI performance rather than outcome correctness. Unlike outcome-graded reliance, which evaluates decisions retrospectively, strategy-graded reliance prospectively looks at cognitive strategies that users might use to maximize overall performance. This shift from focusing on outcomes to strategies offers a more dynamic view of reliance, considering human and AI interactions before the final decision is made. Building on strategy-graded reliance, Guo et al. (2024) advanced the understanding of reliance by applying decision theory to assess reliance on AI based on expected comparative performance. Their measure separates mis-reliance (over- or under-reliance on AI) and discrimination loss (the inability to identify when AI advice is better than human judgment), offering a more refined measurement. The reliance level is quantified as the probability of humans following AI advice if AI's recommendations differ from human judgment, providing insights into how users balance their own judgments with AI suggestions (Guo et al., 2024).

While these measurements offer valuable insights into human reliance levels on AI, they still fall short of capturing the intricacies of problem-solving. Strategy-graded reliance assumes that users can accurately estimate AI's performance, which does not account for cognitive biases or task complexity. Additionally, the measure presupposes a shared understanding between humans and AI, which is often absent in complex tasks. In real-world problem-solving, reliance behaviors might involve subtle cognitive shifts or dynamic adjustments that these models cannot fully capture. Therefore, while useful in structured decision-making, existing measures do not fully address the complexity of human-AI problem-solving, where reliance behaviors are implicit and context-dependent.

### The Current Study

The complexities of problem-solving call for a more comprehensive measure that can account for the diverse reliance behaviors that arise in the context (Joksimovic et al., 2023). Problem-solving tasks, unlike simpler decision-making tasks, require more than binary responses to AI suggestions. They involve exploration, iterative adjustments, and creative reasoning, where reliance behaviors may not always be explicit or immediately observable. In such settings, humans may adopt AI suggestions implicitly, gradually integrating them into their cognitive strategies (Vasconcelos et al., 2023).

Given this gap, this study aims to develop and validate a self-report questionnaire to quantify reliance behaviors during human-AI collaboration in problem-solving. The rationale behind this approach is that self-reported data can capture aspects of reliance behaviors that are difficult to observe directly. Potentially, this measure will provide insights into how students interact with Generative AI during PBL, helping educators better understand the balance between human reasoning and AI assistance.

## Methods and Materials

### Questionnaire Items Development

The study developed questionnaire items that describe specific behavioral cases of how students utilize Generative AI during human-AI problem-solving. The development process was grounded in a systematic approach that sought to map the behavioral traces of human-AI problem-solving across problem-solving stages (i.e., problem identification, information exploration, and solution generation and presentation) and reliance subjects (i.e., Generative AI, self, or peers).

The research team referred to a study (Authors, 2024) that we conducted in a problemsolving context and with samples similar to the current study to develop the initial items. Specifically, Authors (2024) collected 904 responses from students after the problem-solving tasks on two reflection questions: "How would you describe your interaction with ChatGPT during the maze design task? In terms of contributions, did you feel that you, ChatGPT, or both equally contributed to the design of the maze? Please elaborate". After reviewing these reflective responses, we enumerate the behavioral cases of using Generative AI during the problem-solving tasks (e.g., We discussed how to ask prompting questions so Generative AI can give more appropriate outputs; I asked Generative AI to improve my initial draft/solutions for the problemsolving activities). Accordingly, we developed 23 items in total to capture possible reliance behaviors during human-AI problem-solving as shown in Table 1. For example, item R3 (For each activity I did in this class, I spent most of the time interacting with Generative AI) was developed because some students spent considerable time testing the capacity of Generative AI but ignored the problem-solving objectives. Similarly, item R1 (I copied the task descriptions or problem statements to Generative AI for help) was developed because the researchers observed that many students started the task by copying the task description to GPT.

As shown in Table 1, we mapped these questionnaire items to the three main problemsolving stages and the possible subjects that an individual can rely on (i.e., AI, self, and peers). Specifically, during problem identification, AI reliance includes tasks like copying prompts directly into the Generative AI, while self-reliance involves formulating custom prompts, and peer reliance focuses on collaborative clarification of goals or strategies. In the information exploration stage, AI reliance includes extensive interaction and direct use of AI outputs, selfreliance includes critical evaluation and revision of outputs, and peer reliance includes seeking feedback and clarifications from peers. Finally, during solution generation and presentation, AI reliance involves minimal modification of AI outputs, self-reliance reflects independent task completion or rejecting AI suggestions, and peer reliance involves group revisions of AIgenerated content. All the questions were constructed on a five-point Likert-type scale to measure the frequency of the behaviors during human-AI problem-solving (1 = Almost never, 5 = Almost always).

After the initial questionnaire was developed, we invited a panel of experts (2 professors with expertise in human-AI collaboration and 3 graduate students who observed the students' interaction with Generative AI during the study, Authors, 2024) to further review the items to ensure that the items were easy to interpret and choices were easy to make. We improved the wording of the questions to make them clear and easy to interpret. For example, one item, "I reiterated the task requirements to GPT multiple times to let GPT generate good outputs.", was reworded to "I revised my prompting questions so Generative AI can give more appropriate outputs" to demonstrate students' autonomy in using Generative AI. In addition, two items (i.e., R6 and R8) were flagged at this stage because they did not describe specific behaviors, but all 23 items were administered to students after the two problem-solving activities, respectively.

**Table 1.**

*Questionnaire items mapped to the problem-solving stages and reliance categories*

| Problem-solving stages | Reliance categories | | |
| --- | --- | --- | --- |
| | AI reliance | Self-reliance | Peer reliance |
| Problem identification | R1: I copied the task descriptions or problem statements to Generative AI for help. R6: I feel that Gen AI is more capable of doing these activities than me. | R5: I used Generative AI only when I (we) had limited or no clue how the problem could be solved. R9: I designed my own prompts for Generative AI so it could give me the output I need. R15: I revised my prompting questions so Generative AI can give more appropriate outputs. | R20: We discussed how to ask prompting questions so Generative AI can give more appropriate outputs. R21: I asked my peer(s) questions during the problem-solving activities. |
| Information exploration | R3: For each activity I did in this class, I spent most of the time interacting with Generative AI. R8: I feel that Generative AI played a more important role in our problem-solving activities than me. | R7: I asked Generative AI to improve my initial draft/solutions for the problem-solving activities. R10: I read Generative AI's output critically when the output was generated. R11: I modified Generative AI's output to make it more applicable to the problem-solving activities. R12: I spotted errors in the Generative AI's output. R13: I found Generative AI's output is not perfect. | R17: We discussed the Generative AI's output when the output was generated. R22: My peers answered my questions and supported my understanding. |
| Solution generation and presentation | R2: I copied the Generative AI's output as part of our solution. R4: I adopted Generative AI's output without major changes. | R14: I discarded Generative AI's output. R16: I finished a subtask in problem-solving activities with no or limited help from the Generative AI. | R19: My peers helped me modify Generative AI's output. R18: My peer(s) helped me spot errors in the Generative AI's output. R23: I adopted a solution initially proposed by peer(s). |

**Participants**

The participants of this study were first-year and second-year students enrolled in a digital literacy course at an Asian university. In total 3,030 students across various schools in the university enrolled in the digital literacy course, and they were divided into 70 tutorials, with around 20 to 50 students in each. In each tutorial, 5 to 6 students from different schools were purposefully assigned to a group that would work on various problem-solving activities in class. We collected 1818 and 1406 responses in two consecutive weeks towards the end of the semester. The number of responses collected from the second week was lower because some of the instructors did not allow students to use Generative AI that week.

The project received approval from the university's ethics review board to waive consent forms from the participants because the course took place in a regular educational setting, the risk was minimal, the participation was voluntary, and the survey did not take much time from participants and could be considered as their reflections after the problem-solving activities. As such, all students enrolled in this course were encouraged to complete the questionnaire, and no personal information other than their student ID was collected to minimize the risks involved. Based on open data from the university, of all the undergraduate students enrolled in 2023, 54.6% are male, and 45.4% are female; of students enrolled in 2022, 54.2% are male and 45.8% are female.

**Problem-solving Activities**

As mentioned earlier, students in this course participated in various in-class problemsolving activities to improve their digital literacy, with two activities involving Generative AI. During the first activity (3D Maze week), students co-designed and co-solved 3D mazes on a game-based programming platform called 3D Maze. In the first part, they created challenging mazes by placing environmental assets (e.g., rocks, trees, rivers) on a map. In the second part, they solved the maze using pseudo-codes like "turn left" and "if-else loops." For both activities, students are encouraged to use Generative AI to design challenging mazes and solve them with pseudo-codes (See Appendix A for details).

In the second activity (Policy Drafting week), students drafted a policy proposal addressing the fair and ethical use of Generative AI in undergraduate education. They reviewed university guidelines, brainstormed common misuse cases, and evaluated their proposed policies from interdisciplinary perspectives. Students were allowed to use Generative AI freely throughout the problem-solving activity (See Appendix A for details).

**Procedures**

The flipped classroom approach was applied in this course. Before class, students were required to watch instructional videos on related topics and complete the subsequent quizzes. Then, they dedicated two hours per week to complete the two in-class problem-solving activities involving Generative AI. When coming into the class, all the students reviewed key concepts with the instructors. Then, the instructors gave them handouts that included problem-solving activity requirements and emphasized the need for interdisciplinary collaboration among group members. After the problem-solving activities, the instructors reminded students to finish the reliance behaviors questionnaire. However, not all the enrolled students finished the problemsolving activities either because they were absent, time was used up, or their respective instructors adopted different designs rather than using the standard materials provided by the curriculum committee. Some students did not complete the questionnaire.

**Data Analyses**

The data analyses include three steps to ensure that the instrument is reliable and valid for measuring reliance behaviors during human-AI problem-solving. First, during data cleaning, we addressed the issue of careless responses by identifying and removing responses that exhibited patterns indicative of careless responses (see details in the Careless Response below). Following the data cleaning, we conducted an Exploratory Factor Analysis (EFA) using part of the responses from the 3D Maze week to explore the dimensions that emerged naturally from the data to answer what dimensions of reliance behaviors emerge (RQ1). Furthermore, to examine the questionnaire's ability across problem-solving tasks involving Generative AI (RQ2), we conducted two Confirmatory Factor Analyses (CFAs) to validate the factor structure identified in the EFA and assess the model's reliability and validity across two problem-solving tasks: 3D Maze and Policy Drafting tasks. Also, we conducted qualitative validation (Ross et al., 2012; Torrance, 2012) by examining the alignment between students' self-reported questionnaire responses and their actual behaviors using Generative AI, as reflected in their chat histories. Specifically, we analyzed eight chat histories submitted by student groups from two problemsolving activities. We purposefully chose groups that had average questionnaire scores exceeding 4.5 for one of the four factors over the two weeks. We focused on high-scoring groups because their behaviors can be more representative of the factors and can provide explicit and clear evidence of the distinctions among different factors. Lastly, to investigate how undergraduate students' reliance behaviors on Generative AI differ across problem-solving tasks (RQ3), we conducted a descriptive analysis and one-tailed t-test of students' reliance behaviors across the two problem-solving activities to examine undergraduate students' reliance patterns on Generative AI.

### ***Careless Responses***

After an initial review of the data collected, we found some cases of careless responses—answers provided without proper attention or thought, often characterized by random, repetitive, or inconsistent patterns (Kam, 2019). These responses can compromise the quality and accuracy of the data, weaken the relationships among survey variables, and impact factor loading and structure (Meade & Craig, 2012). Therefore, we first screened all the responses to exclude careless responses. For each response, we computed the longstring index, which is the maximum number of consecutive identical responses, to examine the invariability. There is no unified recommended threshold on the longstring index, as participants can respond with identical answers for similar items (Kupffer et al., 2024). Hence, we plotted the distribution of the average length of consecutive identical responses for students' responses to the two problem-solving activities separately. As the reflection surveys administered at the end of both weeks' problemsolving activities included some other items other than the reliance behaviors items, we calculated the longstring index of all the survey questions. For the 3D Maze week (39 items in the reflection survey), the average longstring index ranges from 1.11 to 39, with a mean of 3.75 and a standard deviation of 3.98 (See Figure 1a). For the Policy Drafting week (28 items), the average longstring index ranges from 1.08 to 28, with a mean of 3.84 and a standard deviation of 4.23 (See Figure 1b). We identified outliers, which are data points that exceed 1.5 times the interquartile range above the third quartile of the average longstring index and excluded these responses from further analyses. In the end, 1540 surveys were retained from the 3D Maze week, and 1173 surveys were retained from the Policy Drafting week.

**Figure 1.**

*The Boxplots of the Average Longstring Index of Survey Responses*

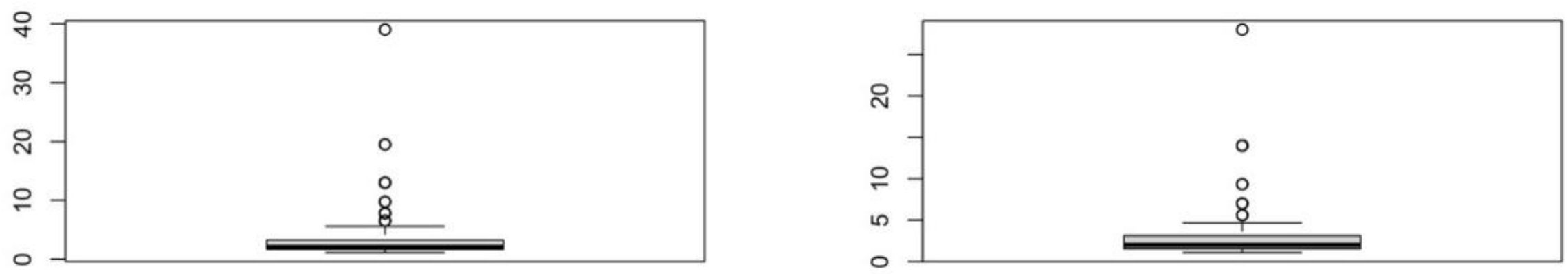

(a) 3D Maze week (b) Policy drafting week

*Notes*. The longstring index was calculated based on complete reflection surveys for both weeks, with 39 and 28 items, respectively.

***The Exploratory Factor Analysis and Confirmatory Factor Analysis***

We first used almost half of the data from the 3D Maze week (800 survey responses) for an EFA. After a preliminary review of correlation among all the factors, the two flagged questionnaire items (i.e., R6 and R8) were weakly correlated with all other items and, therefore, were excluded in the following analyses (See Figure 2 for the correlation visualization). EFA was used to uncover the underlying factor structure that emerges naturally from the data as it can determine the number of latent factors in the factor of interest. This study performed EFA with the Maximum Likelihood (ML) factoring method and oblique rotation. The ML method is suitable for determining latent factors (Fabrigar et al., 1999), and oblique rotation can identify correlated latent factors. Oblique rotation is preferred when developing a new scale (Schmitt & Sass, 2011). This study used a .30 cut-off point for evaluating factor loading (Cudeck & O'Dell, 1994). An item was defined as a cross-loading item and was removed if its factor loading was greater than .3 on two factors and the differences between the loadings were less than .15 (Schmitt & Sass, 2011).

After the EFA, to validate the structure of the factor models, we conducted two CFAs with the remaining 740 survey responses from the 3D Maze week and 1173 survey responses from the Policy Drafting week. Multiple indices could be used for evaluating the final factor model: the comparative fit index (CFI), the root mean square error of approximation (RMSEA), and the standard root mean residuals (SRMR; Kline, 2005). The CFI is a goodness-of-fit measure that compares the fit of the factor model to a null model, accounting for sample size and model complexity. A CFI value over .90 is considered acceptable (Hu & Bentler, 1999). The RMSEA is a fit index to assess how well a hypothesized model, with its estimated parameters, approximates the population covariance matrix. An RMSEA value below .08 is considered acceptable (Marsh et al., 2004). The SRMR is a fit index to measure the average discrepancy between the observed correlation and the model's predicted correlations. An SRMR value below .08 is considered evidence of a good fit (Marsh et al., 2004). To assess the internal reliability of the developed scale, we calculated Cronbach's alpha, with .70 or higher indicating adequate reliability and .6 or higher indicating acceptable reliability (e.g., Azmi et al., 2024). We also calculated the mean inter-item correlations (MICs), with .20 to .40 indicating the optimal range (Piedmont, 2014).

## Results

### RQ1: What Dimensions of Undergraduate Students' Reliance Behaviors Emerge from Question Items?

A total sample of 800 was used for the EFA to determine the latent structure of reliance behaviors during human-AI problem-solving. Figure 2 illustrates the visualization of the correlation matrix among questionnaire items. We adopted polychoric correlation because the scale is ordinal on a Likert scale from 1 to 5 (Holgado–Tello et al., 2010). Then, we performed the Bartlett test of Sphericity and concluded that the variables were correlated enough to proceed with factor analysis ($X^2$ (df = 190) = 6760.16, $p < .05$). We also performed the Kaiser-MeyerOlkin test to assess the adequacy of the sample size for factor analysis and concluded that the factor analysis is appropriate with an overall RMSA value of .92. To determine the number of latent factors, we performed a parallel analysis in R studio using *psych* package (Henson & Roberts, 2006). The parallel analysis suggested 4 factors.

Table 2 illustrates the specific statistics of the 4-factor model. R5 was removed as the factor loading was lower than .30 on all four factors. R7, R17 and R20 cross-loaded on two factors, and the absolute difference between the factor loading was less than .15. All four factors equally include 4 question items each, totaling 16 items on the scale.

**Figure 2.**

*The Correlation Matrix of Questionnaire Items*

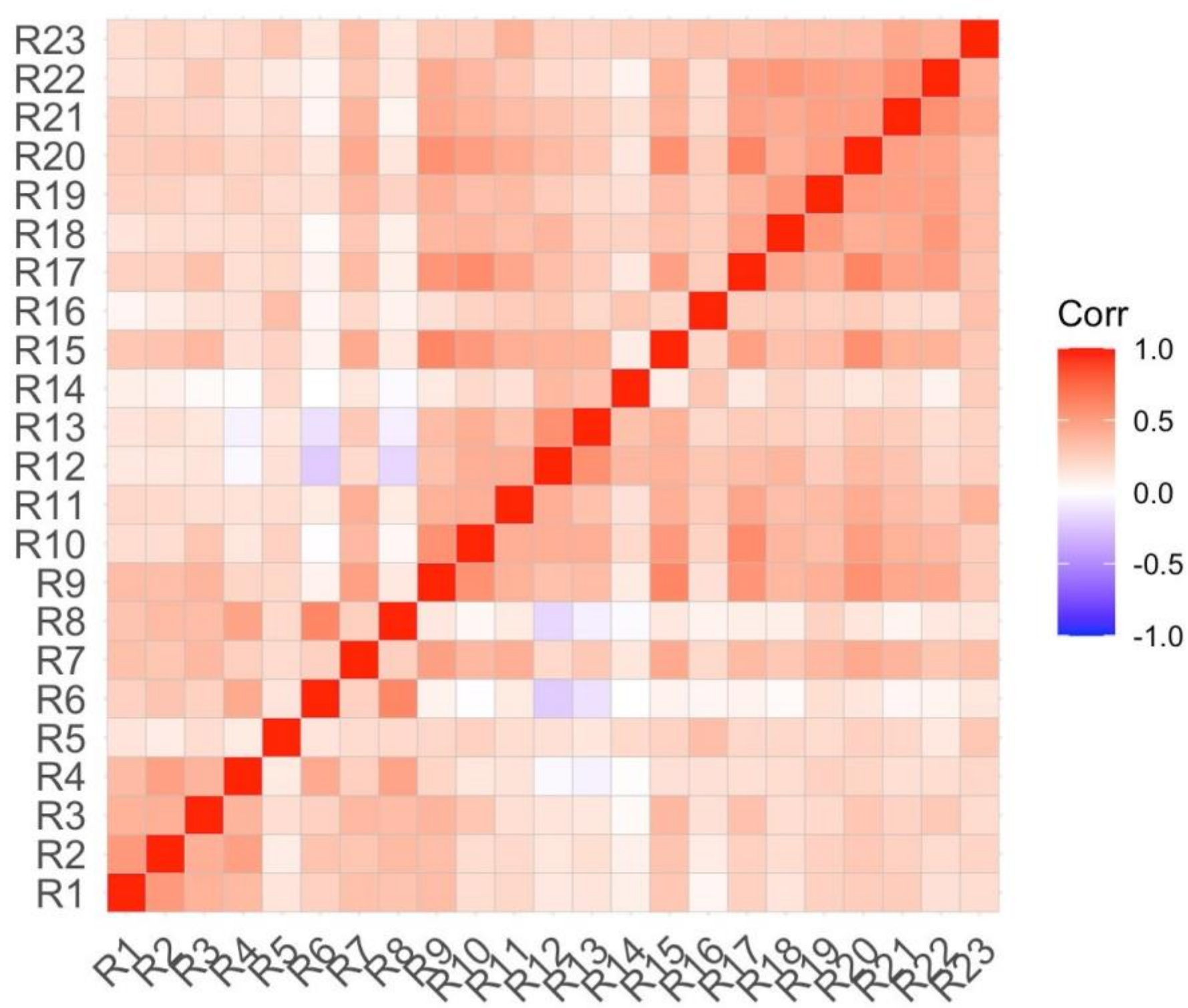

**Table 2.**

*Exploratory Factor Analysis Results*

| | Factor Loading | | | | | |
|---|---|---|---|---|---|---|
| | Factor 1 | Factor 2 | Factor 3 | Factor 4 | Communality ($H^2$) | Uniqueness ($U^2$) |
| R1 | -.16 | .11 | **.68** | .02 | .44 | .55 |
| R2 | -.15 | .03 | **.79** | .03 | .55 | .44 |
| R3 | -.04 | .25 | **.50** | -.08 | .38 | .62 |
| R4 | .14 | -.17 | **.70** | -.18 | .45 | .54 |
| R5 | .16 | -.03 | .11 | .20 | .13 | .86 |
| R7 | .14 | **.32** | **.26** | .04 | .37 | .62 |
| R9 | .11 | **.67** | .12 | -.02 | .63 | .36 |
| R10 | .12 | **.59** | -.10 | .19 | .54 | .46 |
| R11 | .19 | .27 | .02 | .26 | .36 | .63 |
| R12 | -.11 | .25 | -.10 | **.74** | .60 | .39 |
| R13 | -.26 | **.35** | -.05 | **.67** | .51 | .48 |
| R14 | .00 | -.20 | .06 | **.62** | .36 | .66 |
| R15 | .02 | **.65** | .08 | .11 | .57 | .42 |
| R16 | .24 | -.14 | .07 | **.36** | .25 | .74 |
| R17 | **.43** | **.49** | -.08 | -.04 | .56 | .43 |
| R18 | **.67** | .02 | -.09 | .10 | .49 | .51 |
| R19 | **.64** | .04 | .05 | -.01 | .46 | .53 |
| R20 | **.40** | **.47** | .00 | -.02 | .57 | .42 |
| R21 | **.64** | .15 | -.02 | -.03 | .50 | .49 |
| R22 | **.83** | .17 | -.12 | -.24 | .57 | .42 |
| R23 | **.50** | -.15 | .13 | .19 | .38 | .61 |

**RQ2: To What Extent do the Developed Items Demonstrate Reliability across Different Types of Problem-solving Tasks Involving Generative AI?**

We then conducted two separate CFAs with the remaining survey responses collected from the two problem-solving activities to assess whether the scale can be applied to different problem-solving scenarios. With the 3D Maze week data, the fit indices suggested adequate model fit with $X^2(98) = 450.59$, CFI = .89, which is close to .90, RMSEA =.06 < .08, and SRMR = .06 < .08. With the Policy Drafting week data, the fit indices also suggest adequate model fit with $X^2(98) = 604.23$, CFI = .91 > .90, RMSEA =.06 < .08, SRMR = .06 < .08. As the results indicated adequate model fit, we kept all the items in each factor. Factor 1 is associated with questions concerning the revision of prompts or solutions for problem-solving. Thus, it is categorized as *reflective use* (Cronbach's alpha = .74). Factor 2 includes behaviors related to recognizing errors produced by Generative AI and reacting to them. As a result, is named *cautious use* (Cronbach's alpha = .65). Although Cronbach's alpha for this factor is slightly below the conventional threshold of .70, the mean inter-item correlation (MIC) falls within the optimal range of .20 to .40, indicating that the items are sufficiently homogeneous while still reflecting unique aspects of the construct (Piedmont, 2014). Furthermore, removing any item from Factor 2 does not improve the Cronbach's alpha. Thus, all items are retained, as the factor contributes critical conceptual and theoretical value despite its slightly lower reliability. Factor 3 includes behaviors such as copying and pasting and relying on Generative AI for problemsolving, leading us to identify it as *thoughtless use* (Cronbach's alpha = .72). Factor 4 encompasses behaviors related to peer communication and collaboration, so is designated as *collaborative use* (Cronbach's alpha = .76). The Cronbach's alpha of the overall scale is .84.
Table 3 presents the finalized scale with Spearman correlation among each variable.

Moreover, to validate whether the questionnaire items accurately captured reliance patterns, we cross-referenced participants' self-reported responses with their actual interaction patterns in the chat histories. Participants scoring high on reflective use demonstrated autonomy in adapting Generative AI to their needs. For instance, one group critically evaluated GPT's output and provided follow-up prompts like, "I think you misunderstood my request," while another requested revisions to enhance a maze design: "Help me make the following maze more challenging." Fewer groups had high cautious use, and these groups typically identified AI errors or engaged minimally with the tool, suggesting they preferred independent problem-solving. Those scoring high on thoughtless use often copied example prompts from course materials directly into the chat, with some even not replacing placeholders in the example prompts.

However, validating collaborative use from chat histories was more challenging, as such behaviors often involved offline peer interactions. This qualitative validation supports the scale's ability to measure reliance behaviors in human-AI problem-solving.

**RQ3: How Do Undergraduate Students' Reliance Behaviors on Generative AI Differ Across Problem-solving Tasks?**

We then calculated the descriptive statistics, Spearman correlations, and Welch twosample t-tests to explore undergraduate students' reliance behaviors across the two problemsolving activities. As shown in Table 4, reflective use was the highest among the four factors, whereas cautious use was the least frequent among students. Regarding correlations among these factors, a strong correlation between collaborative use and reflective use was found, which might suggest that reflective uses of Generative AI and peer collaborations might reinforce each other. Surprisingly, thoughtless use is moderately related to reflective use, suggesting that students who generally engage in reflective practices may also adopt some thoughtless behavior during human-AI collaboration.

The t-test results revealed significant differences across the two problem-solving activities for reflective, thoughtless, and collaborative uses. Specifically, reflective use was significantly higher in the 3D Maze activity ($t(2284.3) = -9.95$, $p < .001$; M = 4.01, SD = .59 vs. M = 3.76, SD = .67), as were thoughtless use ($t(2220.2) = -15.26$, $p < .001$; M = 3.61, SD = .69 vs. M = 3.15, SD = .82) and collaborative use ($t(2387.4) = -7.28$, $p < .001$; M = 3.92, SD = .62 vs. M = 3.74, SD = .67) than in the Policy Drafting activity. These findings highlight that reliance behaviors vary across different problem-solving tasks.

**Table 3.**

*The Finalized Scale of Reliance Behaviors during Human-AI Problem-Solving with Descriptive Results*

| Categories | Cronbach's Alpha | MICs | M(SD) | Correlation 1 | 2 | 3 | 4 | Questionnaire items |
|---|---|---|---|---|---|---|---|---|
| 1. Reflective Use | .74 | .41 | 4.01(.58) | - | | | | R9: I designed my own prompts for Generative AI so it could give me the output I needed<br>R10: I read Generative AI's output critically when the output was generated<br>R15: I revised my prompting questions so Generative AI can give more appropriate outputs<br>R7: I asked Generative AI to improve my initial draft/solutions for the problem-solving activities |
| 2. Cautious Use | .65 | .37 | 3.45(.69) | .40* | - | | | R12: I spotted errors in the Generative AI's output<br>R13: I found Generative AI's output is not perfect<br>R14: I discarded Generative AI's output<br>R16: I finished a subtask in problem-solving activities with no or limited help from Generative AI |
| 3. Thoughtless Use | .72 | .30 | 3.63(.68) | .28* | .16* | - | | R1: I copied the task descriptions or problem statements to Generative AI for help<br>R2: I copied the Generative AI's output as part of our solution<br>R3: For each activity I did in this class, I spent most of the time interacting with Generative AI<br>R4: I adopted Generative AI's output without major changes |
| 4. Collaborative Use | .76 | .42 | 3.93(.61) | .53* | .37* | .31* | - | R18: My peer(s) helped me spot error in the Generative AI's output<br>R19: My peer(s) helped me modify Generative AI's output<br>R21: I asked my peer(s) questions during the problemsolving activities<br>R22: My peers answered my questions and supported my understanding |

**p* <.001

**Table 4.**

*The Difference in Undergraduate Students' Reliance Behaviors in the Two Problem-Solving Activities*

| | 3D maze week | Policy Drafting week | |
|---|---|---|---|
| Categories | M(SD) | M(SD) | t-test |
| 1. Reflective Use | 4.01(.59) | 3.77(.67) | -9.95* |
| 2. Cautious Use | 3.46(.69) | 3.46(.65) | .19 |
| 3. Thoughtless Use | 3.61(.69) | 3.15(.82) | -15.26* |
| 4. Collaborative Use | 3.92(.62) | 3.74(.67) | -7.27* |

*$p < .001$

## Discussion

The primary rationale for developing the Reliance Behaviors Scale is to uncover how undergraduate students use Generative AI during problem-solving activities and the extent to which they rely on it. With Generative AI becoming increasingly integrated into educational environments and its advanced computational capabilities aiding teams in solving complex issues, such a scale is urgently needed. Previous measures of reliance on AI predominantly depend on decision-making accuracy. In contrast, the current instrument focuses on behaviors during human-AI collaboration without directly linking them to problem-solving outcomes and allows for measuring reliance behaviors in different activities.

Through EFA and CFAs, we verified that the scale includes four factors, each representing a different type of reliance behavior in problem-solving activities: reflective use, cautious use, thoughtless use, and collaborative use. Each factor encompasses specific behaviors that illustrate how students engage with AI during problem-solving tasks. Reflective use includes behaviors where students critically engage with Generative AI. They think about how to frame their questions to get the desired output, critically read and revise AI-generated outputs, and use AI to enhance their initial drafts or solutions. This reflective use suggests a deeper interaction with AI, focusing on improving and refining the problem-solving process. Reflective uses of Generative AI might be associated with reflective thinking, which involves self-regulation and continuous evaluation of one's strategies and outcomes (Gürol, 2011). Reflective thinking encourages individuals to think critically about their strategies, evaluate their progress, and take appropriate steps in problem-solving (Gencel & Saracaloglu, 2018). Reflective thinking is positively associated with cognitive presence and academic achievement (Kim & Lim, 2019).

Similarly, engaging with AI in a reflective manner encourages students to control and regulate their problem-solving processes rigorously, which might foster deeper cognitive involvement and higher academic performance. Future research can examine the correlations between students' reflective use of Generative AI and their problem-solving performance and other types of academic performance.

Cautious use captures behaviors where students recognize and react to errors produced by Generative AI. They spot inaccuracies, acknowledge the imperfection of AI outputs, and sometimes discard AI suggestions. This cautious use indicates a healthy skepticism and a critical evaluation of AI contributions. AI hallucinations—instances where AI produces content that diverges from the input or contradicts established facts—are serious concerns when incorporating Generative AI in educational settings (Zhang et al., 2023). Research shows that LLMs frequently exhibit overconfidence with seemingly authoritative content (Kim et al., 2024). However, if students are aware of the hallucinations and overconfidence, they are more likely to recognize that even the most fluent and convincing responses can be incorrect. Thus, the cautious use of Generative AI can be seen as behavioral traces of users' awareness of hallucinations inherent in Generative AI. This awareness encourages a more vigilant approach, where students actively verify AI-generated information to avoid potential errors and maintain critical engagement with AI outputs.

Thoughtless use describes behaviors where students rely heavily on AI without much critical engagement. They might copy and paste task descriptions to Generative AI or AIgenerated output directly into their solutions, indicating a passive reliance on AI. Previous research has frequently labeled behaviors that accept AI suggestions with no or limited considerations as overreliance. This passive dependence can inhibit students from conducting thorough research and developing their own insights (Kasneci et al., 2023). Li and Little (2023) argued that AI users, regardless of their subject matter expertise, are susceptible to overreliance. At the same time, relying on Generative AI for routine and basic tasks during problem-solving, such as checking grammar issues, or articulating ideas concisely and formally can be beneficial and efficient (Zhai et al., 2024). Hence, it is worth further empirical investigation to determine how the thoughtless use of Generative AI is associated with problem-solving performances and outcomes of other types of learning tasks.

Collaborative use involves behaviors related to peer communication and collaboration in conjunction with Generative AI use, such as students collaboratively spotting errors, modifying AI output, and supporting each other during problem-solving activities. This collaborative approach highlights the social dimension of learning with AI. Collaborative learning is common in higher education (Laal & Laal, 2012). Generative AI can facilitate this by acting as a learning partner, helping students to engage in meaningful dialogues, share diverse perspectives, and improve the collective knowledge of the group (Joksimovic et al., 2023). Students can use AI to explore ideas in their personal space, bring these insights into a shared learning environment, and collectively assess and refine these ideas (Tan et al., 2022). Thus, it is essential that students are capable of collaborating with their group mates, as well as Generative AI tools in the collaborative process. How students interact with their peers and Generative AI in the learning process to optimize the learning process, outcomes, and experiences is a complex issue and warrants further research.

Furthermore, we conducted two CFAs on data collected from two distinct problemsolving tasks. PBL involves diverse and authentic problem-solving activities based on specific learning objectives and encourages collaborative efforts (Wijnia et al., 2019). Thus, to accommodate various problem-solving tasks in undergraduate classrooms, we applied CFAs to two tasks: one focused on programming games and the other on policy drafting. Both analyses showed a good model fit, indicating the scale's potential applicability across various PBL contexts. Together, these categories and items provide a comprehensive measure of reliance behaviors, capturing the nature and frequency of students' interactions with Generative AI across tasks.

The Reliance Behavior Scale focuses on behaviors. We acknowledge that a particular behavior may be beneficial in some situations and potentially detrimental in others. As indicated by the descriptive analysis of reliance behaviors across the two problem-solving activities (Please see Table 4), it is evident that most of the correlations between the factors are moderate or strong, except between thoughtless use and cautious use, which are conceptually distinct. The results imply that people may demonstrate different reliance behaviors in different stages of problem-solving. For instance, reflective behaviors might be highly desirable in contexts requiring deep understanding and critical thinking, while in routine tasks, thoughtless use behaviors might not significantly impact the learning outcomes. By avoiding labeling each behavior as inherently good or bad, the scale provides a flexible tool for examining the diverse ways students use Generative AI across various stages of different learning tasks.

Furthermore, collaborative, reflective, and thoughtless uses are significantly higher in 3D Maze week compared to Policy Drafting week, which highlights the nuanced, context-dependent nature of reliance behaviors. The difference in reliance behavior between Policy Drafting week and 3D Maze week might be explained by cognitive and emotional factors, such as students' confidence in problem-solving, prior exposure to AI tools, perceived ease of use, or attitudes toward collaboration (Van Dongen et al., 2013). While this was beyond the scope of the current study, it might be promising to examine how reliance behaviors vary across tasks requiring different levels of cognitive load, and other related cognitive or emotional factors.

**Limitations**

Several limitations of this study need to be acknowledged. First, the scale was developed and validated with a relatively homogeneous sample of undergraduate students from a single Asian university, which may limit its generalizability. Cultural and environmental factors could influence how students perceive and use Generative AI, potentially impacting reliance behaviors. For instance, cultures with higher levels of technological adaptation may foster greater familiarity and comfort with AI, while economically disadvantaged regions may exhibit lower AI literacy due to limited access to technology and resources (Pinski & Benlian, 2024). Furthermore, the current study's findings may not be directly generalized to K–12 education. Generative AI tools might not yet be widely adopted among younger students due to ethical concerns and limited access. However, adapting the scale to K–12 contexts could offer insights into the reliance behaviors of students at different educational levels. Future research should include a more diverse sample, encompassing students from different educational levels, cultural backgrounds, and learning environments, to broaden the applicability of the findings.

Second, the scale focuses specifically on behaviors related to Generative AI tools, such as those that generate text, images, or code. While Generative AI is a rapidly expanding field, the scale does not account for reliance behaviors with other types of AI, such as intelligent tutoring systems, recommendation systems, or adaptive learning technologies. This limits the scale's applicability across all AI tools. Further research should expand the scale to include behaviors associated with other AI technologies to provide a more comprehensive understanding of human-

AI collaboration.

Third, a notable limitation of this study is that, after removing some items based on EFA results, certain categories of the problem-solving stages were represented by a single question in the questionnaire. For instance, AI reliance for the problem identification stage only includes "I copied the task descriptions or problem statements to Generative AI for help", and peer reliance for the information exploration stage was limited to the single question "My peers answered my questions and supported my understanding". This reduction of questionnaire items may insufficiently capture the complexity of these problem-solving stages. At the same time, while the current questionnaire yielded a good overall Cronbach's alpha, one of the dimensions, Cautious Use, is slightly below the commonly used benchmark (.65 < .70). Thus, further refinement of the survey items is necessary. Future research should expand the number of items in each subcategory to provide more comprehensive coverage of these dimensions and improve the instrument's ability to capture nuanced students' behaviors using Generative AI while collaborating with peers for problem-solving tasks.

Finally, while we had conducted a qualitative validation to compare self-reported behaviors with behaviors observable in the chat histories, such validation was limited to highscoring groups and did not systematically address the full range of self-reported responses. Thus, future studies could adopt a more comprehensive approach, integrating purposeful sampling and analyzing a broader range of responses to further validate and refine the scale.

**Implications**

Developing the Reliance Behaviors Scale has several important implications for subsequent research in human-AI collaboration and educational practice. It enables educators to measure students' reliance on Generative AI tools unobtrusively after problem-solving activities, providing authentic data on how students use AI and can help identify potential issues in students' use behaviors, such as high levels of thoughtless use. Also, educators can leverage the scale results to encourage critical thinking and productive reliance behaviors in problem-based learning contexts. For example, instructors could use insights from the scale to adjust teaching methods by tailoring problem-solving tasks that encourage balanced and critical reliance on both AI and human collaboration. Additionally, the scale can help researchers link specific reliance behaviors to learning outcomes and 21st-century skills, such as problem-solving performance, critical thinking and reflective thinking skills, and communication among students and AI. This information can guide the design of AI tools and instruction that encourage productive use behaviors while reducing overreliance.

# Appendix

## Appendix A

*Problem-solving Activity Requirement and Sample Prompts for Students*

| | | Requirement | Sample Prompts for Generative AI |
|---|---|---|---|
| 3D Maze week | Part 1: Maze Creation | Your group will design a maze in 3D Maze that meets the following design requirements (also stated in the Lesson Overview above):<br>• The width and the length of the maze should be at least 10 units.<br>• The maze should contain all environmental assets (green tree, cherry blossom, rock, block A, block B, river, bridge)<br>• Try your best to make the maze challenging but engaging.<br>Optional: the maze can contain two gems (gems need to be collectible in maze solving) | *I have created a 10 x10 solvable maze. it includes the following elements:*<br>• *1 x starting point represented by "S"*<br>• *1 x end point represented by "E"*<br>• *Walkable paths represented by "0"*<br>• *Obstacles represented by "1"*<br>• *Monsters represented by "M"*<br>• *Gem represented by "G"*<br>• *Heart that must be collected represented by "H"  The following is the maze that I created in a table format. You are an expert on maze design. Can you redesign the following maze to be more challenging by including more elements, and have the path difficult to navigate?* |

| | | |
|---|---|---|
| Part 2: Maze Solving | Your group will solve the maze you created. You can decide how and when you want to use Gen AI, e.g., to generate the pseudo-codes or correct your pseudo-codes. Again you can experiment with different Gen AI tools and decide which one to use. Here is an example prompt that you can use to help you solve the maze. | *In this programming game, I need to solve the maze using pseudo-code to learn programming language. You are a programming teacher. I will tell you in natural language how to navigate the maze, and you need to present me with the correct pseudo codes to solve to maze.*<br>*The pseudo-codes that are available:*<br>*Five actions:*<br>*1. Turn left.*<br>*2. Turn right.*<br>*3. Move forward.*<br>*4. Turn back*<br>*5. attack*<br>*Four conditions:*<br>*1. can move "forward/backward/left/right"*<br>*2. not reach destination*<br>*3. repeat < #*<br>*4. Enemy in front  Three action blocks:*<br>*1. while loop*<br>*2. if else loop*<br>*3. if loop*<br>*Do you understand?* |

| | | | |
|---|---|---|---|
| Policy Drafting week | - | The university has released its position on using Gen AI on campus in 2023. After a semester's practice, the Teaching, Learning, and Pedagogy Division at the university is having an open call for suggestions about the fair and safe use of Generative AI (Gen AI) during undergraduate learning activities. The Division is concerned with Gen AI misuse such as students' excessive use of and over-reliance on AI. They aspire to foster students' critical thinking, creativity, and agency while allowing them to explore latest AI innovations. As students, the focus of teaching and learning activities, your observations and experiences of how Gen AI can be (mis)used, and your suggestions on how to minimize such misuse across campus are critical.<br><br>You and your teammates are undergraduate representatives elected from different schools to | *"The team is working on a problem-solving activity. They need to give one suggestion to the university about how to use Generative AI in their assignments while minimizing misuse. You are an instructor at the university. You need to ask 3 Socratic questions to help them think deeply and evaluate their plan critically. Socratic questions are defined as questions that help learners distinguish things they know from things they don't understand. Some sample Socratic questions are "What is the central aim of the task in this line of thought?"; "How can people from a different viewpoint react to the response?"; and "On what information are you basing that comment?\\\ The team's preliminary suggestion is: \\\""* |

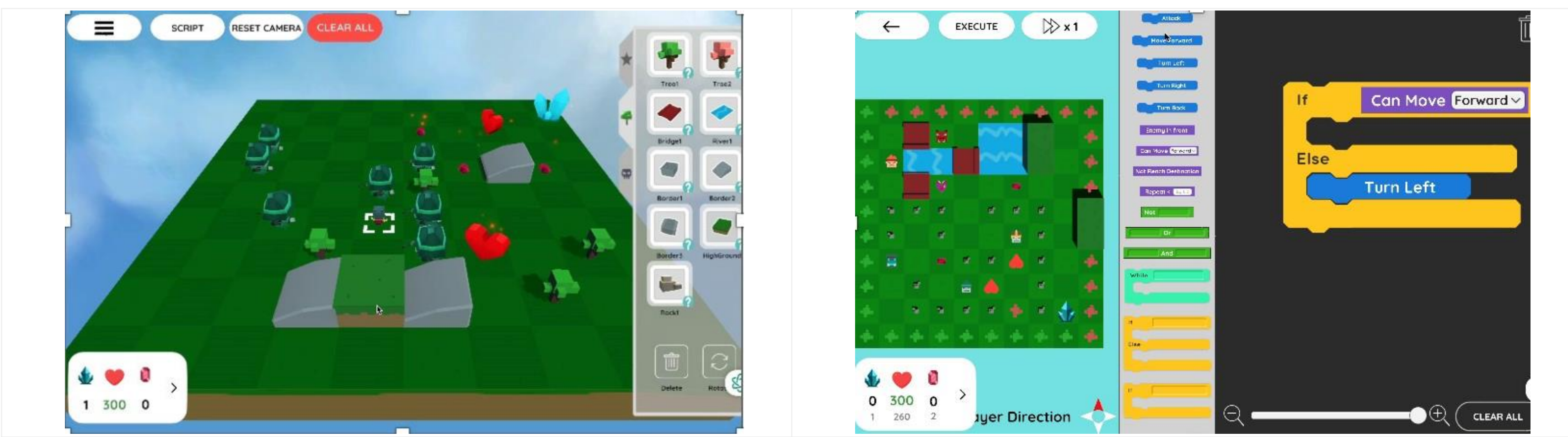
SCRIPT
RESET CAMERA
CLEAR ALL
1 300 0
EXECUTE
x 1
If
Can Move
Forward
Else
Turn Left
CLEAR ALL

| (a) Maze Creation | (b) Maze Solving |
| --- | --- |

**provide specific suggestions for the stakeholders (e.g., students themselves, the university, and instructors) to promote ethical use and minimize misuse of Gen AI on campus**. When considering the use of Gen AI among undergraduate students, you can think about the learning activities you usually have in your respective schools. When your group has developed the suggestion, you should also critically evaluate the suggestion from the perspectives of your respective schools.

**Appendix B**

*An Illustration of the 3D Maze Platform.*